\documentstyle[12pt]{article}
\begin{document}
%
\def\vec#1{\mbox{\boldmath $#1$}}
\def\bold#1{\mbox{\boldmath $#1$}}
\def\quote#1{[#1]}

\newcounter{fignumber}
\newcommand{\capsize}{\small}
\newcommand{\capstyle}{\it}
\newcommand{\capfont}{\small\it}
\newcommand{\mytablecaption}[2]%
{ \refstepcounter{table}%
\centerline{
 {\capsize\bf Tab.~\thetable.}{ }{\capfont #1#2}    
}
  \addcontentsline{lot}{table}                      
  {\protect\numberline{\thetable}{\ignorespaces #1}}%
}
%
\newcommand{\rb}{]}          
\newcommand{\lb}{[}
\newcommand{\Begineq}{\begin{equation}}
\newcommand{\Endeq}{\end{equation}}
\newcommand{\Begineqs}{\begin{eqnarray}}
\newcommand{\Endeqs}{\end{eqnarray}}
\newcommand{\st}{~\hbox{\rm s.t.}~}
\newcommand{\p}{^\prime}
\newcommand{\Oval}[2]{\oval(#1) [#2]}
\newcommand{\bk}{\newline}
\def\endpage{\newpage}
\newcommand{\Item}[1]{\item [#1]}        
\newtheorem{theorem}{Theorem}[section]
\newtheorem{maintheorem}[theorem]{Main Theorem}
\newtheorem{conjecture}[theorem]{Conjecture}
\newtheorem{definition}[theorem]{Definition}
\newtheorem{defandnot}[theorem]{Definition and Notation}
\newtheorem{conventions}[theorem]{Conventions}
\newtheorem{proposition}[theorem]{Proposition}
\newtheorem{mainproposition}[theorem]{Main Proposition}
\newtheorem{lemma}[theorem]{Lemma}
\newtheorem{mainlemma}[theorem]{Main Lemma}
\newtheorem{remark}[theorem]{Remark}
\newtheorem{corollary}[theorem]{Corollary}
\newtheorem{example}[theorem]{Example}

\def\proofmarker{\newline$\Box$\par\noindent}
\newenvironment{Proof}{{\it Proof}\bk\rm}{\proofmarker}
\newenvironment{proof}[1]{{\it Proof#1}\bk\rm}{\proofmarker}

\font\mathy=msbm10 at 12.00pt     
\font\mathys=msbm10               
\font\mathx=msam10 at 12.00pt
\font\mathxs=msam10
\font\mathbf=cmmib10 at 12.00pt
\font\kursivindex= cmsy8
\font\kursivs=cmsy10
\font\kursiv=cmsy10  at 12.00pt
\font\frak = eufm10  at 12.00pt
\font\kursivmed=cmsy10 at 12.00pt
\font\mathmed=cmmi10 at 12.00pt
\font\bigdun = cmdunh10 at 24.88truept
\font\meddun = cmdunh10 at 17.28truept
\font\medsdun = cmdunh10 at 12.00pt
\def\footnotestyle{\footnotesize\rm}
\def\bigtilde{{\kursiv \char'030}}
\def\subsetneq{\hbox{\mathy\symbol{'040}}}
\def\notexists{\hbox{\mathy\symbol{'100}}}
\def\R{\hbox{\mathy R}}
\def\C{\hbox{\mathy C}}
\def\Z{\hbox{\mathy Z}}
\def\N{\hbox{\mathy N}}
\def\Q{\hbox{\mathy Q}}
\def\r{\hbox{\mathys R}}
\def\z{\hbox{\mathys Z}}
\def\n{\hbox{\mathys N}}
\def\nnull{{\hbox{\mathys N}_{_0}}}
\def\c{\hbox{\mathys C}}
\def\uz{^{\hbox{\mathys Z}}}
\def\kur#1{\hbox{\kursiv #1}}
\def\laeq{\hbox{\mathx .}}
\def\hence{\>\hbox{\mathx\symbol{'040}}\>}
\def\circarrow{\>\hbox{\mathy\symbol{'171}}\>}
\def\Rightstar{\stackrel{*}{\Rightarrow}}
\def\emptystring{\hbox{\$}}
\def\circleq{\hbox{$\bigcirc\hspace{-0.82em}
\raisebox{.2ex}{$\scriptstyle \leq$}\>$}}
\def\circl{\hbox{$\bigcirc\hspace{-0.82em}
\raisebox{.2ex}{$\scriptstyle <$}\>$}}
\def\circg{\hbox{$\bigcirc\hspace{-0.82em}
\raisebox{.2ex}{$\scriptstyle >$}\>$}}
\def\circeq{\hbox{$\bigcirc\hspace{-0.82em}
\raisebox{.2ex}{$\scriptstyle =$}\>$}}
\def\disjunion{\stackrel{.}{\cup}}
\def\bigdisjunion{\stackrel{.}{\bigcup}}
\def\limitarrow{\stackrel{\scriptscriptstyle i\rightarrow\infty}
{\rightarrow}}
\def\card{\hbox{\rm card}}
\def\Abb{\hbox{\rm Abb}}
\def\const{\hbox{\rm const}}
\def\dist{\hbox{\rm dist}}
\def\Per{\hbox{\rm Per}}
\def\SPer{\hbox{\rm $\sigma$-Per}}
\def\per{\hbox{\rm per}}
\def\iper{\hbox{\scriptsize\rm per}}
\def\Homocl{\hbox{\rm Homocl}}
\def\id{\hbox{\rm id}}
\def\tr{\hbox{\rm tr}}
\def\area{\hbox{\rm area}}
\def\sgn{\hbox{\rm sgn}}
\def\sgni{\hbox{\scriptsize\rm sgn}}
\def\scat{\hbox{\scriptsize\rm scat}}
\def\start{\hbox{\scriptsize\rm start}}
\def\pmax{p_{\max}}
\def\ymax{y_{\max}}
\def\pimax{\pi_{\max}}
\def\top{\hbox{\scriptsize\rm top}}
\def\Htop{\hbox{H}_{\top}}
\def\lgp{{\lg^\prime}}
\def\up{^\prime}

\def\Tf{T_{k,\pmax}}                   
\def\Tc{T_k}                           
\def\Lf{\Lambda_{k,\pmax}}             
\def\Lc{\Lambda_k}                     
\def\hmap{h}                            
\def\hmapinv{h^{-1}}                    
\def\hf{h_{k,\pmax}}                    
\def\Sf{\Sigma_{k,\pmax}}              
\def\Sc{\Sigma}
\def\SA{\Sigma_M\{1..6\}}
\def\SAA{\Sigma_M(\AA)}
\def\tildeScrit{\tilde{\Sigma}_{\piinfty}}
\def\Spart#1{\Sigma^{\{\pi#1\piinfty\}}}    
\def\Ss{\Sigma_{\AA_0}}
\def\Sa{\Sigma_{\AA}}
\def\Sb{\Sigma_{\BB}}
\def\Snew{\Sigma(\{-1,0,1\})}
\def\Sold{\Sigma(\{1,2,3\})}
\def\Sfull{\Sigma_M(\{1\dots 6\})}
\def\Snpi{\Sigma_{\pi^n}}
\def\Sntildepi{\Sigma_{\tilde\pi^n}}
\def\Snu{\Sigma_\nu}
\def\Smu{\Sigma_\mu}
\def\Snus{\Sigma_\nu/\sigma}
\def\Smus{\Sigma_\mu/\sigma}
\def\Spimax{\Sigma_{\pimax}}
\def\piinfty{{\pi^\infty}}
\def\crit{{\hbox{\scriptsize\rm crit}}}
\def\nucrit{{\nu_\crit}}
\def\mucrit{{\mu_\crit}}
\def\Snucrit{\Sigma_\nucrit}
\def\Scrit{\Sigma_{\piinfty}}
\def\sdag{^\dagger}
\def\sddag{^\ddagger}
\def\Scritdag{\Sigma_{\piinfty}^\dagger}
\def\Scritddag{\Sigma_{\piinfty}^{\ddagger}}
\def\Scrittildedag{\tilde{\Sigma}_{\piinfty}^\dagger}
\def\Gcrit{\G_{\piinfty}}
\def\Gcriteins{\G^{\piinfty}_{\hbox{\sixrm (1)}}}
\def\xn{^{(n)}}
\def\x#1{^{(#1)}}
\def\iz{{i \in \z}}
\def\iN{{i \in \n}}
\def\izdot{{i\in\zdot}}
\def\set#1#2{\{#1,\dots,#2\}}
\def\adic#1#2{\langle #2 \rangle_{#1}}
\def\s{{\bar\sigma}}
\def\modsig{/\sigma}
\def\l{\lambda}
\def\ladic{{\lambda^{-1}}}
\def\minus{\hbox{\rm -}}                  
\def\ones{\hbox{\rm -}11}                 
\def\mones{1\hbox{\rm -}1}
\def\mone{\hbox{\rm -}1}
\def\mtwo{\hbox{\rm -}2}
\def\Ai#1{A_{\hbox{\scriptsize\rm #1}}}
\def\Adh{\hbox{\rm Adh}}
\def\Cyl{\hbox{\rm Cyl}}
\def\Zdot{\dot {\hbox{\mathy Z}}}
\def\zdot{\dot{\hbox{\mathys Z}}}
\def\aa{\hbox{\bf a}}
\def\bb{\hbox{\bf b}}
\def\uu{\hbox{\bf u}}
\def\C{\hbox{\kursiv C}}
\def\D{\hbox{\kursiv D}}
\def\F{\hbox{\kursiv F}}
\def\G{\hbox{\kursiv G}}
\def\I{\hbox{\kursiv I}}
\def\L{\hbox{\kursiv L}}
\def\Lgoth{\hbox{\frak L}}
\def\Ltrio{\hbox{\bf L}}
\def\relL{\sim_{\hbox{\kursivs L}}}
\def\P{\hbox{\kursiv P}}
\def\Pindex{\hbox{\kursivindex P}}
\def\T{\hbox{\kursiv T}}
\def\V{\hbox{\kursiv V}}
\def\Y{\Upsilon}
\def\rel{\hbox{\mathxs r}}
\def\AA{{^{^\circ}}\!\!\!\!A}
\def\OO#1{{^{\circ}}\!\!\!\!#1}
\def\BB{{^{^\circ}}\!\!\!B}
\def\WW{{^{^\circ}}\!\!W}
\def\ZZ{{^{^\circ}}\!\!\!Z}
\def\BBstar{{^{^{\circ}}}\!\!\!B^*}
\def\AAbf{{^{^\circ}}\!\!I}
\def\AAbfindex{{{^{^{^{\circ}}}}\!\!^I}}
\def\alef{{\scriptsize\aleph}}
%
%
%
\title{Formal Languages in Dynamical Systems}
\author{G. Troll\thanks{
Technische Universit\"at Berlin,
Sonderforschungsbereich 288, FB Mathematik MA 7-2,
Str. d. 17. Juni 136, D-1000 Berlin 12, Germany;~
e-mail : troll@math.tu-berlin.de} \\
}
\date{\small January, 1993 \\ to be published in  Acta Univ.
 Carolinae, Math. et Phys., Vol. 34, 2 (1994)}
\maketitle
\begin{abstract}
We treat here the interrelation between
formal languages and those dynamical systems that can be
described by cellular automata (CA).
There is a well-known injective map which
identifies any CA-invariant subshift with a central formal
language. However, in the special case of a symbolic dynamics, i.e.
where the CA is just the shift map, one gets a stronger result:
the identification map can be extended to a functor
between the categories of symbolic dynamics and formal languages.
This functor additionally maps
topological conjugacies between subshifts to
empty-string-limited generalized sequential machines between languages.
If the periodic points form a dense
set, a case which arises in a commonly used notion
of chaotic dynamics, then an even more natural map to assign
a formal language to a subshift is offered. This map
extends to a functor, too.
The Chomsky hierarchy measuring the complexity of formal
languages can be transferred
via either of these functors
from formal languages to symbolic dynamics and proves to be
a conjugacy invariant there. In this way it acquires a dynamical meaning.
After reviewing some
results of  the complexity of CA-invariant subshifts,
special attention is given to a new kind of invariant
subshift: the trapped set, which originates from
the theory of chaotic scattering and for which one can
study complexity transitions.
\end{abstract}
\section{Introduction}
This article is organized as follows:
Section 2 treats cellular automata
and in particular the shift map as
dynamical systems on  subshifts, i.e. closed and shift invariant
subsets of sequence spaces
(cf. \cite{CALosAlamos83,CALosAlamos89}).
Besides some standard CA-invariant
subshifts a new kind (the trapped set)
is defined, which originates from chaotic scattering theory
To this purpose formal multiscattering systems (FMS) are introduced
and two examples, the truncated double horseshoe and
the truncated sawtooth are given (cf. \cite{Troll91phd,Troll92pre}).
Section 3 is a short introduction in formal
language theory (cf. \cite{Salomaa73,Ginsburg75,HopcroftUllm90}).
In particular, we define a strucure of families of
languages called trio whose closure properties
have a dynamical correspondence.
An important example for trios are the levels of the
Chomsky hierarchy, which offers
 a means to classify formal
languages according to their structural complexity.
The morphisms which will make
a link to symbolic dynamics are empty-string-limited
generalized sequential machines (GSM). They preserve trios.
Section 4 addresses this problem of how to link the two categories
of subshifts and languages (cf. \cite{CulikHurd89b,Troll92pre}).
If one restricts oneself to shift dynamics then it is possible
to construct an injective functor between these categories.
In this case empty-string-limited GMS correspond to
conjugacies between subshifts.
Finally, in section 5, the complexity of dynamically
generated languages is discussed. Some general results
(cf. \cite{CulikHurd89b}) for CA
and some applications in formal multiscattering systems
(cf.\cite{Troll91phd,Troll92pre}) are reviewed.
Another approach to the complexity of dynamical systems
which uses chaotic limits and factors of finite automata
instead of finite partitions is presented in \cite{Kurka93pre}.
\section{Dynamical Systems on Subshifts}
\label{Ssymdyn}
\subsection{Basic definitions}
\label{SSdef}
We discuss here dynamical systems whose dynamical map
operates on sequences. We permit sequences over some finite set
$\AA$ of abstract symbols with cardinality $\#\AA>1$
and define the {\it full shift} over $\AA$
as the complete set of sequences $\AA\uz$.
\par
Take the usual topology on $\AA\uz$, i.e. the product topology of
the discrete topology on $\AA$.
A base for this topology is given by the cylinder sets
which consist of all sequences having a common middle segment
around the zero index:
\Begineq
\Cyl(a_1\dots a_n) =\{s\in\AA\uz;~s_i=a_{i+[{n-1\over 2}]+1}~\hbox{for}
{}~-[{n-1\over 2}]\leq i\leq [{n \over 2}]\}
\Endeq
which are both open and closed.
This topology is compact (by Tychonov) and metrizable
by the {\it sequence metric} $d$:
\Begineq d(a,b) =
\sum_{\iz}{\bar{\delta}_{a_i b_i} \over 2^{|i|} }
   \label{Emetric}\Endeq
where $a,b \in \AA\uz$, $\bar{\delta}_{xy} =0$
if $x=y$, $1$ otherwise.
\par
Among the continuous maps operating on
$\AA\uz$ shift maps and cellular automata (CA)
are of special interest because they are compatible
with the group structure of the index set $\Z$:
\par
Denote by $\sigma_{\AA}$ the (right) {\it shift map} on $\AA\uz$:
$(\sigma(a))_i =a_{i+1}$. Usually we write just $\sigma$ for
$\sigma_{\AA}$.
A {\it subshift} is a closed $\sigma$-invariant set in $\AA\uz$.
$\Sigma$ is a {\it finite subshift} if it is determined by a finite
set $F$ of forbidden symbol strings, which are not permitted
to appear in sequences.
We call $\Sigma$ {\it cyclic} if the periodic (w.r.t. $\sigma$) sequences
are dense.
A dynamical system given by s subshift $\Sigma\subset\AA\uz$
and the shift map $\sigma|\Sigma\rightarrow\Sigma$ is
called a {\it symbolic dynamics}.
A natural generalization leads to {\it cellular automata (CA)}.
A (1-dimensional) CA is any continuous map $\tau$ on $\AA\uz$
which commutes with the shift map.
Since any CA is automatically uniformly continuous
an equivalent description of a CA
(cf. \cite{Hedlund68,Hedlund69}) can be given by a local function
$f_\tau\,:~\AA^{2R+1}\rightarrow\AA$
with the property $(\tau(s))_i=f_\tau(s_{i-R},\dots,s_{i+R})$ for
any $s\in\AA\uz$.
\par
The appropriate morphisms for topological dynamics are
{\it semi-conjugacies}. We define them here just for shift maps
as any continuous map $\Phi$ between subshifts
$\Sigma_{\AA},~\Sigma_{\BB}$ which
intertwines between the respective shift maps
$\sigma_{\AA},~\sigma_{\BB}$:
\Begineq
\Phi\,:~\Sigma_{\AA} \rightarrow \Sigma_{\BB},\quad
\sigma_{\BB}\circ\Phi =\Phi\circ\sigma_{\AA}
\Endeq
If, moreover, $\Phi$ is a homeomorphism, it is called a {\it conjugacy}
(between shift maps). It may be interpreted as a continuous change
of variables which identifies the shift maps on $\Sigma_{\AA}$ and
 $\Sigma_{\BB}$.
\subsection{Some standard CA-invariant subshifts}
\label{SSinvsubsh}
As for general dynamical systems one introduces the
forward (and backward) limit set,
the periodic sets and the non-wandering set.
Because of the commutability of a CA with the shift map they
are also subshifts. Let $\tau$ be again a CA over $\AA\uz$ which
need not necessarily be invertible.
\begin{definition} \rm
\begin{itemize}
\item The forward limit set $\Lambda^+(\tau)$ is the intersection
of all forward images:
\Begineq
\Lambda^+(\tau) =\bigcap_{n=0}^\infty \tau^n(\AA\uz)
\Endeq
\item The periodic set $\Pi(\tau)$ is the topological closure of
the set of all cycles:
\Begineq
\Pi(\tau) =\overline{\Per(\tau,\AA\uz)}
\Endeq
\item The non-wandering set $\Omega(\tau)$:
\Begineq
\Omega(\tau) =\{s\in\AA\uz;~\forall~ \hbox{\rm neigbourhood of}~s\,
\exists\,n\in\N\st \tau^n(U)\cap U\not = \emptyset\}
\Endeq
\end{itemize}
\end{definition}
\begin{remark}\rm
\begin{enumerate}
\item All the sets just introduced are $\tau$-invariant subshifts.
\item The forward limit set is the maximal invariant subshift of $\AA\uz$.
\item $\Pi(\tau)\subset\Omega(\tau)\subset\Lambda^+(\tau)$.
\item The dynamics on these invariant sets can be very complicated.
We recall here the phenomenon of chaos, for which we shall assume
the following (topological) definition \cite{Devaney86}: \\
As invariant set we take a $\tau$-invariant subshift
$\Sigma\subset\AA\uz$. Following \cite{Devaney86} we call
$\tau|\Sigma$ {\it chaotic} if
\begin{itemize}
\item{(i)} $\tau$ depends sensitively on initial conditions. \\
($\exists\,\delta>0\,\forall\,s\in\Sigma\,
\forall\,\hbox{neighb.}U_s\subset\Sigma\,\exists\,t\in U_s,\,n\in\N\st
d(\tau^n(s),\tau^n(t))>\delta$)
\item{(ii)} $\tau$ has a dense orbit.
\item{(iii)}  $\Sigma$ is $\tau$-cyclic, i.e. $\Pi(\tau|\Sigma)=\Sigma$.
\end{itemize}
\end{enumerate}
In \cite{BrooksCair92} it is shown (in the general context of
dynamical systems on metric spaces) that condition (i) can be replaced
by requiring the invariant set to be infinite.
\end{remark}
\subsection{Another CA-invariant subshift: the trapped set}
\label{SStrapped}
We are going to introduce now a generalization of the
usually studied limit sets of a CA:
\begin{definition}\rm
Let $\Gamma$ be a closed subset of the compact sequence space $\AA\uz$
and $\tau~:~\AA\uz\rightarrow\AA\uz$ a CA.
Furthermore, we require that either $\tau$ be the shift map or
 $\Gamma$ be a subshift.
We define as trapped set $\Lambda(\tau,\Gamma)$ the
$\Gamma$-maximal subshift
\Begineq
\Sigma:=\Lambda(\tau,\Gamma) =\bigcap_{\iz}\tau^i(\Gamma)
\Endeq
\end{definition}
This definition is motivated by scattering systems
(cf. next subsection).
The selection principle by which certain
sequences in $\AA\uz$ and consequently
(because of compactness) certain segments are forbidden
in $\Sigma$ is often called pruning.
\subsubsection{Trapping in Formal Multiscattering Systems (FMS)}
\label{SSfms}
\begin{definition} \rm
Let $M$ be a metrizable space and $T\,:\>M\rightarrow M$ be a
bijective map. Furthermore, let $Q\subset M$ be a compact subset
and define the following images:
\Begineq
I^i :=T^i(Q)\cap Q,\quad i\in\Z
\Endeq
The pair $(T,Q)$ is called a {\it formal multiscattering system} (FMS)
if the following conditions are satisfied:
\begin{itemize}
\item[(i)] {\bf homeomorphism property:}
The restriction $T|Q \rightarrow T(Q)$ is a homeomorphism.
\item[(ii)] {\bf escape property:}
$\forall\,n\in\N\,:~T^{n+1}(I^{-n}) \setminus Q\not = \emptyset$.
\item[(iii)] {\bf trapping property:}
$\forall\,n\in\N\,:~T^{n+1}(I^{-n}) \cap Q\not = \emptyset$.
\item[(iv)] {\bf no-return property:}
if $x \in Q,~T(x)\not\in Q$ then $\forall\,n\in\N\,:~T^{n}(x)\not\in Q$.
\end{itemize}
\end{definition}
\begin{remark} \rm
\begin{enumerate}
\item The set $Q$ may be interpreted as the interaction set of some
finite range potential. The dynamics restricted to the
complement $M\setminus Q$ is then the comparison dynamics of some
scattering problem.
\item Properties (iii) -- (iv) are also valid for $T^{-1}$ because
of the bijectivity of $T$ and the following lemma.
\item Properties (iii), (iv) are equivalent to
$\forall \,n\in\N\,:~I^n\not =\emptyset$ and $T(I^{n})\setminus Q\not=Q$.
\end{enumerate}\
\end{remark}
\begin{lemma}
\label{Lnested}
With the notation of the preceding definition we have:
\begin{itemize}
\item[(i)] $I^{n+1}=T(I^n)\cap Q= \bigcap_{i=0}^{n+1}I^i$
\item[(ii)] $T^{-j}(I^i)=I^{-j}\cap I^{i-j}$, for $i\geq j>0$.
\end{itemize}
\end{lemma}
\begin{definition} \rm
We define the trapped set of a FMS $(T,Q)$ as its
$Q$-maximal invariant set, i.e.
\Begineq
 \Lambda:=\Lambda(T,Q):=\bigcap_{i\in\Z}T^i(Q)
\Endeq
\end{definition}
\begin{lemma}
The trapped set $\Lambda$ of a FMS is non-empty and compact.
\end{lemma}
\begin{Proof}
$\Lambda\not =\emptyset$ follows from the trapping property (iii)
which implies by lemma \ref{Lnested} the finite intersection
property $\bigcap_{i=-n}^{n}T^i(Q)=\bigcap_{i=-n}^n I^i = I^n \cap I^{-n}
\not = \emptyset$. Compactness of $Q$ yields the assertion.
\end{Proof}
\begin{definition} \rm
\begin{itemize}
\item[(i)] A FMS $(T,Q)$ is called {\it expansive} if
$T|\Lambda$ is expansive in the usual sense, i.e.
if $\exists\, \delta>0\st$ $\{x\not = y$ implies
$\exists\, n\in\Z\,:~d(T^nx,T^ny)>\delta\}$.
\item[(ii)] A finite open cover $\alpha$ of $\Lambda(T,Q)$ is called a
generator for the FMS $(T,Q)$ if $\forall$
sequences $(A_i)\in\alpha^{\z}:$ $\bigcap T^i(\bar{A_i})$
 contains no more than 1 point in $Q$.
If $\bigcap T^i({A_i})$ contains no more than 1 point, $\alpha$ is
called a weak generator.
\end{itemize}
\end{definition}
As in the standard situation (cf. \cite{Walters82}) one has
\begin{lemma}
The FMS is  expansive iff it has a generator iff
it has a weak generator.
\end{lemma}
For an expansive FMS the existence of a symbolic dynamics
is shown by practically copying the standard proof \cite{Walters82}.
\begin{theorem}
If $(T,Q)$ is an expansive FMS, then there is a surjective
semi-conjugacy $\Phi:\,\Sigma\rightarrow\Lambda(T,Q)$ from
a subshift $\Sigma$ over some finite alphabet $\AA$.
\end{theorem}
If there is a cover of disjoint sets one finds the following
strengthening:
\begin{corollary}
If the FMS $(T,Q)$ has a disjoint generator, then $\Phi$ can be
chosen to be a topological conjugacy.
In this case the trapped set $\Lambda(T,Q)$ is totally disconnected.
\end{corollary}
Finally we are introducing parameterized families of FMS.
\begin{definition}\rm
\label{DtruncFMS}
A {\it truncated family of FMS} is given by a parameterized family of
FMS $(T_\kappa,Q_\kappa)_{\kappa\in J}$ over some interval $J\subset\R^n$
as parameter set, where each $Q_\kappa$ is the truncation of a common
set $Q\supset Q_\kappa$ by a family of level lines given
 by a cut-off function
$ f:M\rightarrow \R$ and an evaluation function $e:J\rightarrow \R^+$:
\Begineq
Q_\kappa = \{u\in Q\,:~|f(u)|\leq e(\kappa)\}
\Endeq
\end{definition}
We shall now give a simple example:
\subsubsection{Example 1: The truncated double horseshoe}
\label{SStrunchorse}
Consider the following variant of Smale's piecewise linear horseshoe
map:
As usual, we perform on the unit square
$Q_1=[0,1]^2$ first a linear horizontal contraction and
a vertical expansion by positive factors $\lambda<1$ and
$\lambda^{-1}>1$ respectively, followed by
a folding, so that the folded parts fall outside $Q_1$.
 But we choose a double folding%
\addtocounter{fignumber}{1}%
\newcounter{FSmale}%
\setcounter{FSmale}{\value{fignumber}}%
, whose purpose
is to make the horseshoe map $T=T^{(H)}$
restricted to $I^{-1}:= Q_1\cap T^{-1}(Q_1)$
not only linear
but to remove the reflection contained in Smale's horseshoe map,
i.e.
\Begineq
 DT|I^{-1} = \pmatrix{\lambda & 0 \cr
                            0      & \lambda^{-1} \cr}
\Endeq
To construct a symbolic dynamics of this hyperbolic system
we use the partition of
$I^{-1}$ into two disjoint horizontal strips $H_0$ and $H_1$ and
define a topological conjugacy%
\footnote{i.e. $h$ is a homeomorphism and $h\circ T=\sigma\circ h$.}
\Begineq
h\,:~ \Lambda(T,Q_1) \rightarrow \{0,1\}\uz,\quad
u \mapsto (a_i)_{\iz}\quad\hbox{with}~T^i(x)\in H_{a_i}
\Endeq
We introduce the following lexical pseudo-order
(not antisymmetric) on the full shift $\{0,1\}\uz$:
Let $a, b \in \{0,1\}\uz$ and define the horizontal line
$H_a=\bigcap_{i=0}^{\infty}H_{a_i}$, then
\Begineq
a < b :\Leftrightarrow
{}~\hbox{for the smallest}~i_0\geq 0\st a_{i_0}\neq b_{i_0}~\hbox{one has}~
a_{i_0}<b_{i_0}
\Endeq
This pseudo-order corresponds to the arrangement of the corresponding
horizontal lines:
\Begineq
\pi_y\left(H_{(a_i)_{i=0}^\infty}\right)
 \geq \pi_y\left(H_{(b_i)_{i=0}^\infty}\right)
\Leftrightarrow a\geq b
\Endeq
where we denoted by $\pi_y$ the projection to the $y$-component.
\par
We now interpret $Q_1$ as the maximal interaction set, which
will be truncated by the pruning function $\pi_y$ to yield
a 1-parameter family of pruned scattering systems
with interaction sets%
\addtocounter{fignumber}{1}%
\newcounter{Ftruncdouble}%
\setcounter{Ftruncdouble}{\value{fignumber}}%
 \Begineq
Q_{\ymax} :=\{u\in Q_1:~ \pi_y(u) \leq \ymax\}
\Endeq
In the notation of definition \ref{DtruncFMS} the evaluation function $e$
is the identity on $[0,1]$ and the cut-off function $f$ is the
projection $\pi(y)$.
The corresponding map $T_{\ymax}$ is defined by
modifying the horseshoe map $T$ only outside the box $Q_{\ymax}$ to make
points drift to infinity without returning whereas
$T_{\ymax}|Q_{\ymax} = T|Q_{\ymax}$.
This yields a 1-parameter formal multiscattering system
$(T_{\ymax},Q_{\ymax})_{\ymax\in(0,1]}$.
Each time a test particle ``jumps" out of the box $Q_{\ymax}$
it escapes for good.
The trapped set we are interested in is the
$Q_{\ymax}$-maximal invariant set
\Begineq
\Lambda(T_{\ymax},Q_{\ymax}) = \{u\in Q_1;
\pi_y(T_1^i(u))\leq \ymax ~\forall\,i\in\Z\}
\Endeq
It consists of all trajectories which never jump out
of the box $Q_{\ymax}$.
\par
Lowering $\ymax$ will make it easier to get out of the box
$Q_{\ymax}$. Formerly trapped trajectories manage to escape,
so that the trapped set is bound to shrink.
Actually, we are going to describe the evolution of the trapped set
in the symbolic dynamics defined above, where the variation
of the selection condition $y\leq\ymax$ will lead to
pruning rules determining which symbol sequences
correspond to trapped trajectories.
\par
In this model truncation is trivially transferred to
 symbolic dynamics
where the trapped set equivalent to
$\Lambda(T_{\ymax},Q_{\ymax})$
is the subshift
\Begineq
 \Sigma_\nu:=\{s\in\{0,1\}\uz;~\forall\,l\in\Z\,:~
(s_i)_{i\geq l}\leq \nu\}
\Endeq
where $\nu$ is the binary expansion of $\ymax$.
\subsubsection{Example 2: The truncated sawtooth}
\label{SStrunctooth}
This example originates from a
physical model
of the scattering of a point particle in an infinite
array of non-overlapping elastic
scatterers which are placed at unit distance from each other along
the $y$-axis (cf. \cite{TrollSmil89,Troll92}).
The symbolic dynamics of this model is a family of FMS
over the alphabet $\AA=\{-1,0,1\}$ with interaction set
\Begineqs
\Gamma_\nu&=&\{s\in\AA\uz;~|f(s)|\leq \nu\},~
\nu\in \BB^{\nnull},\quad \BB =\{-2,-1,0,1,2\} \nonumber\\
 f\,:~\AA\uz&\rightarrow&\BB^{\nnull},~
s \mapsto s_0(s_1+s_{-1})\dots(s_i+s_{-i})\dots
\Endeqs
and lexical ordering on $\BB^{\nnull}$.
\section{Formal Languages}
This section introduced the category of formal languages
and some subcategories.
It is based on \cite{Salomaa73,Ginsburg75}.
\label{Sfl}
\subsection{Basic definitions}
\label{SSfldef}
We start from a finite set $\AA$ of abstract symbols,
which will be called {\it alphabet} in the following.
Define the free semi-group of words or (finite) strings
$\AA^* = \bigcup_{n=0}^{\infty}\AA^n$,
which is called the {\it total language} over $\AA$.
The semi-group operation is the concatenation.
Its unit element is the empty string $\emptystring$.
Powers correspond to repetitions of symbols, the length $\lg(s)$ of a
string $s\in\AA^*$ yields a formal logarithm.
The total language minus the empty string is
written as $\AA^+$.
Any subset $L$ of the total language $\AA^*$ is called
a {\it language}.
A word $x\in L$ is a subword or {\it segment} of a word $y\in L$, written
$x \prec y$ if there are $a,b\in\AA^*\st y=axb$.
A {\it homomorphism} $h$ between total languages is understood with
respect to the semi-group structures.
If, moreover, $h^{-1}(\{\emptystring\})=\{\emptystring\}$,
then $h$ is called $\emptystring${\it -free}.
When we speak of an {\it inverse homomorphism} $h^{-1}$ we
mean the set valued function of preimages of singletons under $h$.
\subsection{Grammars Generating Formal Languages}
\label{ASrewriting}
In this section we shall discuss how to
generate a formal language by a grammar.
This concept is based on the notion of a
rewriting system.
\paragraph{Rewriting systems}
A {\it rewriting system} $RW=(\AA,P)$ is given by an alphabet $\AA$
and a finite set $P\subset\AA^*\times\AA^*$.
The pairs $(S,T)\in P$ are referred to as {\it rewriting rules}
or {\it productions} and written as $S\rightarrow T$.
\par
Assume $S\rightarrow T,~T\rightarrow U\in P$.
We are allowed to apply these productions also to subwords
within a word and using the notation ``$\Rightarrow_{RW}$" or
in short ``$\Rightarrow$" to
generate a new word directly, i.e. in 1 step:
$R_1SR_2 \Rightarrow R_1TR_2$ or in finitely many%
\footnote{The relation $\Rightstar$ is the
reflexive transitive closure of the binary relation $\Rightarrow$
i.e. $\Rightstar:=\bigcup_{i=0}^{\infty}(\Rightarrow)^i$
where $(\Rightarrow)^0:=\{(a,a);~a\in\AA\}$.}%
steps:
$R_1SR_2 \Rightarrow R_1TR_2\Rightarrow R_1UR_2$ which we write as
$R_1SR_2 \Rightstar R_1UR_2$.
\begin{definition} \rm
A {\it generative grammar} $ G=(\AA_T,\AA_V,X_0, P)$ is
now a rewriting system $(\AA, P)$ where the alphabet
$\AA =\AA_T\disjunion\AA_V$ is partitioned
into {\it terminals} and {\it variables} or {\it nonterminals}.
The nonterminal alphabet $\AA_V$ contains
a distinguished letter, namely the {\it initial letter} $X_0$.
In any production $S\rightarrow T\in P$ the
word $S$ being ``processed" must contain
at least one variable, i.e. $S\not\in\AA_T^*$.
\end{definition}
\par
When we use $G$ to generate its associated
language $L_G$ we start with the
initial letter $X_0$, and do not stop
in the generating process before we have produced a terminal word
in $\AA_T^*$:
\Begineq
L_G =\{S\in\AA_T^*:~ X_0\Rightstar S\}
\Endeq
\paragraph{Chomsky hierarchy}
Grammars are classified by imposing restrictions
on the form of productions.
The most common classification is the
{\it Chomsky hierarchy}:
\begin{definition} \rm
A grammar $ G =(\AA_T,\AA_V,X_0, P),\>\AA=\AA_V\disjunion\AA_N$,
is of the {\bf type $\boldmath i$}
if the following restrictions $(i)$ are satisfied for
the rewriting rules ($X,Y\in\AA_V$):
\begin{itemize}
\item[(0)]no restrictions;
\item[(1)]the rewriting rules ''depend on the context", i.e.
they have all the following form:
$QXR\rightarrow QAR$, where $Q,R,A\in\AA^*$ and
$A$ is not the empty string $\emptystring$ with the only possible
exception: if $X_0\rightarrow \emptystring$ appears as a production,
then $X_0$ must not occur on the right side of any production.
\item[(2)]All rewriting rules have the form
$X\rightarrow A$, where $A\in\AA^*$;
\item[(3)]the rewriting rules concatenate only on one side or
replace by a terminal $S\in\AA_T$, i.e. their form is:
$X\rightarrow SY$ or $X\rightarrow S$.
\end{itemize}
\end{definition}
Type 1 grammars restrict type 0 grammars by requiring
that their productions (with the possible exception of
$X_0\rightarrow\emptystring$) are not
length decreasing%
\footnote{This is even an equivalent characterization
of type 1 grammars.}
(the length of a word is the number of its primitive
(i.e. either terminal or nonterminable) symbols).
\par
A {\it language is of type} $\boldmath i$,
$\boldmath i$ $\in\{0,1,2,3\}$,
if it can be generated by a grammar of type $i$. The common names
for type $i$ languages are
{\it recursively
enumerable, context sensitive, context free} and
{\it regular}, respectively. We denote the set of type $i$
 languages by $\Ltrio_i$.
This classification is properly nested, i.e.
$\Ltrio_{i+1}\subset\Ltrio_i$.
A class properly between type 0 and type 1 languages
(type 1/2) are the {\it recursive} languages.
A language $L$ is recursive if both $L$ and its complement
$\AA^*\setminus L$ are recursively enumerable.
The {\it Chomsky complexity} $\chi(L)$ of a language $L$ is defined
as $(-1) \times$ the type $i$ of its simplest generating grammar.
\par
Certain subfamilies of regular languages are of interest, too,
such as {\it finite complement languages},
where a finite set of words $F\subset \AA^*$ are
forbidden segments. They correspond to
finite subshifts.
\par
In fact, besides these language types many other types have been studied.
Many of them have common properties.
We shall define just one underlying structure, called trio structure.
For this we have to introduce families of languages.
\begin{definition} \rm
Let $\AAbf$ be an infinite alphabet and $\Ltrio$
a set of languages over $\AAbf$, i.e. $\Ltrio\in 2^{\AAbfindex^*}$
 The pair $(\AAbf,\Ltrio)$ is called a
{\it family of languages} if
\begin{itemize}
\item[(i)] $\forall\, L\in\Ltrio~\exists\,\AA\subset\AAbf\st$
$\AA$ is finite and $L\subset\AA^*$.
\item[(ii)] $L\neq \emptyset$ for some $L\in\Ltrio$.
\end{itemize}
\end{definition} \rm
A trio is now a family of languages with certain closure properties:
\begin{definition}  \rm
A {\it (full) trio} is a family of languages closed
under $\emptystring$-free (arbitrary) homomorphisms, inverse
homomorphisms, and intersections with regular languages.
\end{definition}
\begin{example} \rm
The families of regular, context free, and recursively enumerable
languages are full trios. The families of context sensitive and
recursive languages are trios that are not full.
\end{example}
\subsection{Acceptors and Morphisms of Languages}
\label{SSautomata}
\paragraph{Acceptors}
Automata are mathematical models of devices that process
information by giving responses to inputs.
Formal language theory views them as scanning devices
or {\it acceptors} able to
recognize, whether a word belongs to a given formal language.
A hierarchy of automata corresponds
to the Chomsky hierarchy of languages, in the sense
that any class i language is recognized by a class i automaton
and conversely any class i automaton produces
a class i language as output.
The corresponding automata are called Turing machines (0)
linear bounded automata (1), pushdown automata (2) and
finite deterministic automaton (3). We refer for details to
the literature (e.g.\cite{Salomaa73,Ginsburg75,HopcroftUllm90}).
\paragraph{Morphisms}
An intrinsic justification of a classification scheme are
closure properties of its classes under naturally associated
morphisms. We choose the
maps induced by $\emptystring$-limited
generalized sequential machines (GSM).
They are generalizations of homomorphisms.
Their special importance for our discussion lies in the fact that they
correspond to the morphisms of subshifts, i.e. their semi-conjugacies
(cf. section \ref{SSfunc}).
\begin{definition} \rm
A  {\it generalized sequential machine (GSM)}
is given by the six-tuple $M=(Z,\AA,\BB,\omega,\tau,z_0)$,
where $Z$ (state set), $\AA$ (input alphabet), and $\BB$ (output alphabet)
are finite sets, $\tau\,:~Z\times \AA\rightarrow Z$
(next state map) and
$\omega:\,Z\times \AA \rightarrow \BB^*$ (output map)
are maps.
$M$ is called $\emptystring$-free if the empty string $\emptystring$
is not an assumed value of $\omega$.
We extend the output map $\omega$ and
the next state map $\tau$ to $Z\times\AA^*$ by
$\tau(z,\emptystring):=z$,
$\omega(z,\emptystring):=\emptystring$, and
for $s\in \AA^*$, $a\in\AA$ recursively
$\tau(z,sa):=\tau(\tau(z,s),a)$,
 $\omega(z,sa):= \omega(z,s)\omega(\tau(z,s),a)$ by concatenation.
\newline
The map
\Begineq
M:\>\AA^* \rightarrow \BB^*, \quad
s \mapsto \omega(z_0,s)
\Endeq
is called a {\it GSM map}.
A GSM M is called {\it $\emptystring$-limited} on a given
language $L\subset \AA^*$ if there is a $k>0$ s.t.
$\forall\,s\in L$ : if $s=xyz$ and
$\omega(\tau(z_0,x),y)=\emptystring$ for some
$x,y,z\in\AA^*$ then $\lg(y)\leq k$.
\end{definition}
An important property of
$\emptystring$-limited GSM maps is
that they keep trios and hence
 the  Chomsky classes invariant if one neglects
the empty string (cf. \cite{Ginsburg75}).
\begin{lemma}
\label{LGSMclosed}
For each trio $\Ltrio$, each $L\in\Ltrio$, and each GSM map
$M$ $\emptystring$-limited on $L$ we have:
\begin{itemize}
\item[(i)] $M(L)\setminus\{\emptystring\}\in\Ltrio$
\item[(ii)] If $\Ltrio$ is a full trio or if
$\Ltrio$ is closed under union with $\emptystring$
or if $a\in L,~M(a)=\emptystring$ implies $a=\emptystring$, then
even $M(L)\in\Ltrio$.
\item[(iii)] $M^{-1}(L)\in\Ltrio$.
\end{itemize}
\end{lemma}
\begin{corollary}
Let $\Lgoth$ be the language category with
languages as objects and $\emptystring$-limited GSM maps
as morphisms. Any full trio forms a subcategory.
\end{corollary}

\section{Subshifts and Languages --}
\vskip -.2ex
\centerline{\large\bf how they fit together}
\medskip
In this section we are going to study how topological dynamics
can be linked with formal language theory.
There are two approaches. In the first,
which we will not treat further, one extends formal language
theory to encompass bi-infinite words. In the second, one
establishes a bijective correspondence between subshifts and
so-called central languages.
 We are going to follow two tracks.
 The first, introduced in \cite{CulikHurd89b}, constructs a bijection
between the set of all subshifts and the set of all central languages,
which will be defined presently. In the second, we restrict our
attention to chaotic symbolic dynamics, which we required in particular
to be cyclic (cf. section \ref{SSinvsubsh}),
and identify those languages which correspond
naturally to a cyclic subshift (cf. \cite{Troll92pre}).
\subsection{Dynamically generated languages}
First we define the languages we associate to general and cyclic
subshifts, respectively. These are the associated {\it central}
language
which consists of all segments of  permitted sequences and
the associated  { \it  cyclic} language, which consists of all full
periods of a cycle, or more formally:
\par\noindent
Define the {\it periodization map}
\Begineq
\label{Eperiodmap}
\zeta_{\AA}\,:~ \AA^*\rightarrow\Per(\sigma,\AA\uz),\quad s
\mapsto \bar s
\Endeq
where $\bar{ s}_i :=s_{i\,\bmod\, \lg(s)}$.
\begin{definition} \rm
\label{Dassoclang}
Let $\Sigma,\Sigma^\prime\subset\AA\uz$ be an arbitrary and
a cyclic subshift (i.e. $\Pi(\sigma,\Sigma)=\Sigma$), respectively. Then
\begin{itemize}
\item[(i)] $\L(\Sigma) = \{a\in\AA^*;~\exists s\in\Sigma\,:\,a\prec s\}$
is called the {\it associated central language} and
\item[(ii)] $\L_c(\Sigma^\prime)
=\zeta_{\AA}^{-1}(\Per(\sigma,\Sigma^\prime))\cup\{\emptystring\}$
is called the {\it associated cyclic language}.
\end{itemize}.
\end{definition}
Observe that $\L_c(\Sigma^\prime)$ contains together with any
word $w$ all its repetitions $w^i$, $i>0$ and all their
cyclic permutations.
An easy exercise shows the following
\begin{lemma}
\label{Llangopinj}
The maps $\L$ and $\L_c$ are injective.
\end{lemma}
Therefore different subshifts can still be distinguished on the
level of their associated languages.
\subsection{Adherences and centers}
A map in the opposite direction, i.e. from
the set of all languages to the set of all subshifts,
is the {\it adherence}.
It maps a language onto the subshift of all sequences whose
segments are also segments of words of the language:
\begin{definition} \rm
Let $L\subset\AA^*$.
\Begineq
\Adh(L) =\{s\in\AA\uz;~\forall a\prec s\exists w\in L\st a\prec w\}
\Endeq
is called the {\it adherence} of the language $L$.
\end{definition}
Obviously, $\Adh(L)$ is closed and shift invariant and hence a subshift.
The language operators $\L$ and $\L_c$ are in fact inverses of
the properly restricted adherence operator $\Adh$.
The right restrictions can be characterized as invariants
of the center operators $\C,\C_c$:
\begin{definition} \rm
We define the following two operations on the set of languages
over $\AA$:
\begin{itemize}
\item[(i)] The {\it center} $\C(L)$ of a language $L$ is the language
\Begineq
\C(L) = \{a\in\AA^*;~\forall N>0\,\exists x,y\in\AA^*,\,
\lg(x),\lg(y)\geq N\st xay\in L\}
\Endeq
A string in $\C(L)$ is sometimes called {\it bi-extensible} in $L$.
\item[(ii)] The {\it cyclic center} $\C_c(L)$ is
\Begineq
\C_c(L) =\{a\in\AA^*;~\forall i>0 \forall\,
\hbox{\rm permutations}~\pi(a^i)\,
\exists x,y\in \AA^*\st x\pi(a^i)y\in L\}
\Endeq
\end{itemize}
\end{definition}
We call a language $L$ {\it central} if $\C(L)=L$ and
{\it cyclic} if $\C_c(L)=L$.
One sees from this definition immediately
that the associated central language of a subshift is central and that
the associated cyclic language of a cyclic subshift is cyclic.
Conversely, by the following theorem linking subshifts and languages,
one sees that any central language
is the associated central language of some subshift, and likewise
any cyclic language is the
associated cyclic language of some cyclic subshift.
\par
We now state the announced theorem, whose
 first part (i) is due to \cite{CulikHurd89b}.
\begin{theorem}
\label{TLAdhisId}
Let $L$ be a language over $\AA$.
\begin{itemize}
\item[(i)] If $\Sigma$ is any subshift over $\AA$, then
\Begineqs
\Adh(\L(\Sigma)) &=& \Sigma  \label{ELAdhisIda}\\
\L(\Adh(L)) &=& \C(L) \label{ELAdhisIdb}
\Endeqs
\item[(ii)] If $\Sigma^\prime$ is a cyclic subshift over $\AA$, then
\Begineqs
\Adh(\L_c(\Sigma^\prime)) &=& \Sigma_c \label{ELAdhisIdc} \\
\L_c(\Adh(L)) &=& \C_c(L) \label{ELAdhisIdd}
\Endeqs
\end{itemize}
\end{theorem}
\begin{Proof}
ad (i) eq. \ref{ELAdhisIda}:
By applying the definitions one sees that
$\L(\Adh(\L(\Sigma)))=\L(\Sigma)$. The injectivity of $\L$ implies
the first assertion. \\
For eq. \ref{ELAdhisIdb}
let first $a\in\L(\Adh(L)$), i.e. $\exists\,s\in\Adh(L)\st
a=(a_1\dots a_n) =(s_1\dots s_n) \prec s$.
Let $N>0$ and $b:=s_{1-N}\dots s_{n+N}$. We have $b\in \L(\Adh(L))$
and $a\prec b\prec s$. By the definition of the adherence
$\exists\,c\in L\st a\prec b \prec c$. Hence $a\in C(L)$.
Conversely, let $a\in C(L)$. Then $a$ must have an infinite
number of bi-extensions in $L$. This implies the existence
of a nested sequence of words $w^{(i)}\in C(L)$, $i\geq 0$,
($w^{(0)}=a$) with strictly increasing length
 $\lg(w^{(i)})\rightarrow\infty$. By compactness of $\AA\uz$ the set
$W=\bigcap_{i=0}^\infty\Cyl(w^{(i)})$ is nonempty, hence a singleton
$\{w^{(\infty)}\}$ which by construction is contained in $\Adh(L)$.
As $a \prec w^{(\infty)}$, we get $a\in\L(\Adh(L))$.
\par\noindent
ad (ii) eq.\ref{ELAdhisIdc}
"$\subset$": Let $s\in\Adh(\L_c(\Sigma^\prime))$. Then
$\forall\,a\prec s\,\exists\,w\in\L_c(\Sigma^\prime)\st a\prec w$.
But $w\in\L_c(\Sigma^\prime)$ means that $\bar w\in\Sigma^\prime$.
As $\Sigma^\prime$ is cyclic
($\overline{\Per(\sigma,\Sigma^\prime)}=\Sigma^\prime$), we conclude
$s\in\Sigma^\prime$.
"$\supset$": As the adherence of a language is a subshift, it is
enough to show $\Per(\sigma,\Sigma^\prime) \subset
\Adh(L_c)(\Sigma^\prime)$. So let $s \in\Per(\sigma,\Sigma^\prime)$.
Then $\exists\,a\in\AA^*\st s=\bar a$ and $\forall\,i\geq 1\,:$
$a^i$ and all its permutations are in $\L_c(\Sigma^\prime)$.
Therefore $s=\bar a\in\Adh(\L_c(\Sigma^\prime))$.
\par\noindent
ad (ii) eq.\ref{ELAdhisIdd}
"$\subset$": Let $a\in\L_c(\Adh(L))$. Then $\bar a \in \Adh(L)$, i.e.
$\forall\,c\prec \bar a\,\exists\,x,y\in\AA^*\st xcy\in L$.
In particular $\forall\, i\geq 1\,\forall$ permutations $\pi(a^i)
\,\exists\, x,y\in\AA^*\st x\pi(a^i)y\in L$.
Hence $a\in\C_c(L)$.
"$\supset$": Let $a\in\C_c(L)$. Then
$\forall\, i\geq 1
\,\exists\, x,y\in\AA^*\st xa^iy\in L$. Hence $\bar a\in\Adh(L)$, so that
$a\in\L_c(\Adh(L))$.
\end{Proof}
\begin{corollary}
\label{Csshiftandcentral}
\begin{itemize}
\item[(i)] The pair of maps $(\L,\Adh)$
give a bijective correspondence between central languages
and subshifts over $\AA$.
\item[(ii )] The pair $(L_c,\Adh)$
gives a bijective correspondence between cyclic languages and
cyclic subshifts over $\AA$.
\end{itemize}
\end{corollary}
\subsection{Functors from the category of languages to the category
of subshifts}
\label{SSfunc}
Let {\frak SD} be the category of symbolic dynamics,
i.e. its objects are the subshifts and
its morphisms the semi-conjugacies of shift maps.
The subcategory of cyclic subshifts will be denoted by {\frak SD}$_c$.
Let {\frak L} be the
category of languages with languages as objects and
$\emptystring$-limited GSM maps as morphisms.
We shall now extend the language operators $\L$ and $\L_c$ to
yield functors.
\par
The key observation is that
analogously to CA, (cf. \cite{Hedlund68,Hedlund69}) there
is a description of semi-conjugacies by local functions.
\par
Let $\Phi:\,\Sa\rightarrow {\Sigma}_{\BB}$
be a semi-conjugacy, first between arbitrary subshifts.
Since $\Sa$ is compact
$\Phi$ is even uniformly continuous. Therefore, for each $l\in\N_0$
there exists a minimal odd number $N(l)\geq 0$, s.t. for the discrepancy
 $i_\delta$ of two sequences $s,t$
(i.e. half the length of the longest common middle segment),
 we find
\Begineq
i_\delta(s,t)\geq N(l) \Rightarrow i_\delta(\Phi(s),\Phi(t))\geq l
\label{ENfunc}
\Endeq
Since $\Phi$ commutes
with the shift map, it bears a close resemblance
with a cellular automaton (CA) but for
the possibility to change the sequence space.
Like a CA, $\Phi$ is determined by a {\it local function}
albeit defined only on a subset $D\subset\AA^{2R+1}$, $2R+1:=N(1)$:
\Begineqs
\phi:\,&D&\rightarrow\BB, \quad
\phi(s_{i-R}\dots s_i\dots s_{i+R})=(\Phi(s))_i \nonumber\\
&D&=\{a\in\AA^{2R+1};~\exists s\in\Sigma_{\AA}\st a\prec s\}
\label{Elocalfunc}
\Endeqs
We call $R$ the {\it radius of uniform continuity} of $\Phi$.
\par
Conversely, let $r\in\N$, $D\subset\AA^{2r+1}$,
$\phi\,:~D\rightarrow\BB$. Denoting by $\Sigma(F)\subset\AA\uz$
the finite subshift generated
by the forbidden set $F=\AA^{2r+1}\setminus D$ we get
a semi-conjugacy $\Phi\,:~\Sigma(F)\rightarrow\Sigma_{\BB}$,
$(\Phi(s))_i=\phi(s_{i-r}\dots s_i\dots s_{i+r})$
onto some finite subshift $\Sigma_{\BB}\subset\BB\uz$.
Thus, we have shown
\begin{lemma}
There is a bijective correspondence between semi-conjugacies
of finite subshifts and local functions.
\end{lemma}
This local representation of $\Phi$ can be used to
prove the following
\begin{lemma}
\label{LGSMrecoding}
Each semi-conjugacy between subshifts
can be simulated by a $\emptystring$-limited GSM mapping both
 on the associated
central languages and on the associated cyclic languages (if the
subshifts are cyclic).
\end{lemma}
\begin{Proof}
Let $\Phi:\,\Sa\rightarrow {\Sigma}_{\BB}$
be a semi-conjugacy between two subshifts $\Sa\subset\AA\uz$ and
${\Sigma}_{\BB}\subset\BB\uz$.
Let $\phi:\,\AA^{2R+1}\supset D\rightarrow\BB$, $R\geq 0$,
be the associated local
function of eq. \ref{Elocalfunc}.
For convenience,
extend $\phi$ in some arbitrary way to the whole $\AA^{2R+1}$.
\par\noindent
(i) First we examine the central language.
This part is an adaptation of a proof given by Culik II et al.
\cite{CulikHurd89b} for cellular automata.
Define a map
\Begineq
\label{Ecentralfunc}
\L(\Phi):\>\L(\Sa)\rightarrow\L(\Sigma_{\BB}) ,~
(s_i)_{i=1}^n \mapsto
\cases{\phi(s_1\dots s_{1+2R})\dots \phi(s_{n-2R}\dots s_n)
&if $n>2R$\cr
        \emptystring &otherwise\cr}
\Endeq
Observe that a string of length $n\geq 2R$ is mapped by $\L(\Phi)$ to
a string of length $n-2R$.
Obviously, $\L(\Phi(\Sa))=\L(\Phi)(\L(\Sa))$.
\newline
Next one defines a GSM,
whose associated GSM map will be just $\L(\Phi)$: \\
Let $M =(Z,\AA,\BB,\omega,\tau,z_0,Z_f)$
with state set $Z=\bigcup_{i=0}^{2R}\AA^i$,
initial state $z_0=\emptystring$,
next state function $\tau$ and output function $\omega$
defined for $z\in Z$, $a\in\AA$, as follows:
\Begineq
  \tau((z_i)_{i=1}^r,a) =\cases{(z_i)_{i=1}^ra &if $r <2R$\cr
                                (z_i)_{i=2}^ra &if $r=2R$\cr} ;~
\omega((z_i)_{i=1}^r,a) =\cases{\emptystring &if $r<2R$\cr
                                   \phi\bigl((z_i)_{i=1}^ra
\bigr) &if $r=2R$\cr}
\Endeq
$M$ is a $\emptystring$-limited GSM
on $\L(\Sa)$ because only strings
shorter than $2R$ are mapped to the empty string $\emptystring$.
Obviously, the GSM map $M|\L(\Sa) =\L(\Phi)$.
\par\noindent
(ii) For the cyclic language we proceed similarly.
Observe, that we identified repetitions of cycles $s^i$, $i>0$,
$s\in\L_c(\Sigma)$.
We define a length conserving $\L_c(\Phi)$ and
periodize $s=(s_i)_{i=1}^n\neq\emptystring$ by $s_{n+i}:=
(\zeta(s))_{n+i}=s_i$:
\Begineq
\label{Ecyclicfunc}
\L_c(\Phi):\>\L_c(\Sa)\rightarrow\L_c(\Sigma_{\BB}) ,~
(s_i)_{i=1}^n \mapsto
\cases{
\phi(s_1\dots s_{1+2R})\dots \phi(s_n\dots s_{n+2R})
 &if $(s_i)_{i=1}^n\neq\emptystring$\cr
\emptystring &otherwise
\cr
}
\Endeq
Again we get
$\L_c(\Phi(\Sa))=\L_c(\Phi)(\L_c(\Sa))$.
Next we concatenate to each $s\in\AA^*$ an end marker $\bullet$,
i.e. we identify $s$ with $s\bullet$.
The $\emptystring$-limited GSM
simulating the semi-conjugacy $\Phi$ is defined by
$Z=\bigcup_{i=0}^{8R^2}\AA^i$, $z_0=\emptystring$,
$\tau(z_0,a):=[a;a]$, $h,t\in\AA^*$.
The string $h$ is going to store
the leading ($2R$ or up to $4R^2$) symbols of an input word $s\bullet$
(see cases 1,2 in the eq. below):
\Begineqs
  \tau(\left[(h_i)_{i=1}^r;(t_i)_{i=1}^r\right],~a)
&=&\cases{[(h_i)_{i=1}^ra;(t_i)_{i=1}^ra]&if $r<2R,~a\neq\bullet$\cr
        [\left((h_i)_{i=1}^r\right)^{2R};\left((t_i)_{i=1}^r\right)^{2R}]
&if $r<2R,~a=\bullet$\cr
          [(h_i)_{i=1}^{2R};(t_i)_{i=2}^{2R}a]
                                          &if $r= 2R,~a\neq\bullet$\cr
           \left[(h_i)_{i=1}^{r};(t_i)_{i=1}^{r}\right]
                                          &if $r\geq 2R,~a=\bullet$\cr
          }
\nonumber \\
\omega(\left[(h_i)_{i=1}^r;(t_i)_{i=1}^r\right]~,a) &=&
\left\{\matrix{
\emptystring \quad \hbox{if}~ r<2R \hfill\cr
       \phi\left((t_i)_{i=1}^{2R}a\right) \quad
 \hbox{if}~ r= 2R,~a\neq\bullet\hfill\cr
       \phi(t_1\dots t_{r}h_1)\dots\phi(t_{r}h_1\dots h_{2R})
                                  ~\hbox{if}~ a=\bullet\hfill\cr
}\right.
\Endeqs
where $\phi(t):=\phi(t_1\dots t_{2R+1})$ if $\lg(t)>2R$.
Again, we get $M|\L_c(\Sa) =\L_c(\Phi)$.
\end{Proof}
\begin{corollary}
Definition \ref{Dassoclang} together with
equations \ref{Ecentralfunc} and \ref{Ecyclicfunc}
defines two functors $\L\,:~\hbox{\frak SD}\rightarrow\hbox{\frak L}$
and $\L_c\,:~\hbox{\frak SD}_c\rightarrow\hbox{\frak L}$.
\end{corollary}
The lemmas \ref{LGSMclosed} and \ref{LGSMrecoding} yield the
announced closure properties:
taking factors does not increase complexity%
\footnote{Compare this to the analogous statement about topological
entropy in \cite{Kurka92pre}.}%
{}.
\begin{theorem}
\label{Tchomskyclosure}
Any semi-conjugacy
$\Phi$ between two subshifts preserves any full trio
 the associated (central or cyclic) language belongs to;
in particular, if $\Sigma$ is an arbitrary subshift,
$\Sigma^\prime$ a cyclic subshift then:
\Begineqs
\chi\left(\dot{\L}_c\left(\Phi(\Sigma^\prime)\right)\right)
&\leq&
\chi(\dot{\L}_c(\Sigma^\prime))   \nonumber\\
\chi\left(\dot{\L}\left(\Phi(\Sigma)\right)\right)
&\leq&
\chi(\dot{\L}(\Sigma))
\Endeqs
where $\dot{L}:=L\setminus\{\emptystring\}$ for any language $L$.
If furthermore, $\Phi$ is a conjugacy then the inequalities
just stated become equalities.
\end{theorem}
\section{Complexity  of dynam. generated languages}
\label{ScxDL}
In this section we first return to general CA.
\subsection{Sofic Systems}
\label{SSsofic}
The simplest way to generate a language
dynamically is by applying some CA a finite number of times
to a full shift.
As it happens these so-called sofic systems
yield just the regular languages:
\begin{definition}
Let $\tau$ be a cellular automaton (CA) over the shift $\AA\uz$
and $\Sigma_M$ a subshift of finite type.
The subshift $\tau(\Sigma_M)$ is called a {\it sofic system}.
\end{definition}
One has (cf. \cite{Hurd88phd}) the following:
\begin{proposition}
A subshift $\Sigma$ is sofic iff $\L(\Sigma)$ is regular.
\end{proposition}
\subsection{Some results for CA and their "standard" invariant sets}
\label{SSresCA}
We quote here some results from \cite{CulikHurd89a}.
First one observes the following constraint for limit languages:
\begin{theorem}
\label{Tlimitlangone}
For every cellular automaton $\tau\,:~\AA\uz\rightarrow\AA\uz$
the complement of the forward limit language
$\AA^*\setminus\L(\Lambda^+(\tau))$ is recursively enumerable.
\end{theorem}
Apart from this constraint any language is realisable by
the limit set of a CA in the following sense:
\begin{theorem}
\label{Tlimitlangtwo}
If $L\subset\AA^*$ has a recursively enumerable complement, then
there exists a CA $\tau\,:~\BB\uz\rightarrow\BB\uz$,
a regular language $R\subset\BB^*$, and a homomorphism
$h\,:~\BB^*\rightarrow\AA^*\st$:
\Begineq
h(\L(\Lambda^+(\tau))\cap R)=L
\Endeq
\end{theorem}
In particular examples exist, for which the the central language
of the forward limit set is not recursively enumerable.
\par
For the periodic set one has the following results:
\begin{theorem}
\label{Tperiodiclang}
\begin{itemize}
\item[(i)] If $\tau$ is any CA, then $\L(\Pi(\tau))$ is
recursively enumerable.
\item[(ii)] If $L\subset\AA^*$ is  recursively enumerable,
then there exists a CA $\tau\,:~\BB\uz\rightarrow\BB\uz$,
a regular language $R\subset\BB^*$, and a homomorphism
$h\,:~\BB^*\rightarrow\AA^*\st$:
\Begineq
h(\L(\Pi(\tau))\cap R)=L
\Endeq
\end{itemize}
\end{theorem}
Theorems \ref{Tlimitlangtwo} and \ref{Tperiodiclang} (ii)
 describe only the complexity
of a sublanguage of a limit language.
They do not yield better lower limits that theorems
\ref{Tlimitlangone} and \ref{Tperiodiclang} (i).
Examples of limit languages with any of the introduced Chomsky
complexities can be found in
\cite{Hurd90a,Hurd90b}. Instead, we shall treat in the next section
examples of families of FMS which illustrate
how a variation in complexity accompanies a transition
to chaos of the associated dynamical system.
\subsection{Results for some simple truncated families of FMS}
\label{resFMS}
For this subsection cf. \cite{Troll91phd,Troll92pre}.
Here we treat the  examples of subsections \ref{SStrunchorse} and
\ref{SStrunctooth}. In these models
one is interested in complexity
transitions arising by variation of the family parameter
and accompanying transitions to chaos of these models.
For other complexity transitions cf. \cite{CrutchfieldYoun89}.
\paragraph{Truncated horseshoe}
The first observation one makes is, that the
parameterization map
\Begineqs
\{0,1\}^{\nnull}\cong [0,2]&\rightarrow& 2^{\AA\uz} \nonumber \\
 \nu&\mapsto&\Sigma_\nu:=\{s\in\{0,1\}\uz;~\forall\,l\in\Z\,:~
(s_i)_{i\geq l}\leq \nu\}
\Endeqs
is locally constant everywhere but on the
set
\Begineq
 \V=\{a\in\{0,1\}^{\n};~\forall\,i\in\Z\,:~\sigma^i(a)\leq a\}
\Endeq
which is called the {\it bifurcation set} of the family of FMS.
One observes the following complexity transition:
\begin{theorem}
Suppose $\nu\in\V\cap\{1,2\}$.
Then both central language $\L(\Sigma_\nu)$ and
cyclic language $\L_c(\Sigma_\nu)$ are regular iff
$\nu\in\Q$.
\end{theorem}
\paragraph{Truncated sawtooth}
This system becomes chaotic at the critical value $\nucrit=1-1\bar 0$.
For $\nu<\nucrit$ the trapped set is finite, hence
the associated languages regular.
At the critical value one finds:
\begin{theorem}
The associated languages $\L(\Sigma_{\nucrit})$ and
$\L_c(\Sigma_{\nucrit})$ are context sensitive but not context free.
\end{theorem}

%
%
%
\section*{Acknowledgement}
\pagestyle{plain} \indent
This paper was presented at the 21st Winter School of the
Charles University, Prague, on Abstract Analysis, Section Topology,
Jan. 23 -30, 1993.
This work was supported by the
Deutsche Forschungsgesellschaft DFG
within a project of the Sonderforschungsbereich 288.
\newcommand{\noopsort}[1]{} \newcommand{\singleletter}[1]{#1}
\ifx\undefined\bysame
\newcommand{\bysame}{\leavevmode\hbox to3em{\hrulefill}\,}
\fi


\begin{thebibliography}{10}

\bibitem{BrooksCair92}
J.~Brooks, G.~Cairns, G.~Davis, and P.Stacey, {\em On {Devaney's} definition of
  chaos}, The American Math. Monthly {\bf 99,4} (1992), 332.

\bibitem{CrutchfieldYoun89}
J.~P. Crutchfield and K.~Young, {\em Computation at the onset of chaos},
  Complexity, Entropy and Physics of Information (Reading, Massachusetts)
  (W.~Zurek, ed.), Addison--Wesley, Reading, Massachusetts, 1989.

\bibitem{Devaney86}
R.~L. Devaney, {\em An introduction to chaotical dynamical systems}, Benjamin
  Cummings Publ. Co. , Inc., Menlo Park, 1986.

\bibitem{CALosAlamos83}
D.~Farmer, T.~Toffoli, and S.~Wolfram (eds.), {\em Cellular automata, proc. of
  an interdisciplinary workshop 1983}, Physica D 10, 1984.

\bibitem{Ginsburg75}
S.~Ginsburg, {\em Algebraic and automata--theoretic properties of formal
  languages}, North--Holland, Amsterdam, 1975.

\bibitem{CALosAlamos89}
H.~Gutowitz (ed.), {\em Cellular automata: Theory and experiment, {Proc.} of
  the {Los Alamos} workshop 1989}, Physica D 45, 1990.

\bibitem{Hedlund68}
G.~A. Hedlund, {\em Transformations commuting with the shift}, Topological
  Dynamics (J.~Auslander and W.~G. Gottschalk, eds.), Benjamin, 1968,
  pp.~259--289.

\bibitem{Hedlund69}
\bysame, {\em Endomorphisms and automorphisms of the shift dynamical system},
  Math. Syst. Theor. {\bf 3} (1969), 320--375.

\bibitem{HopcroftUllm90}
J.~E. Hopcroft and J.~D. Ullmann, {\em Introd. to automata theory, languages,
  and computation}, Addison--Wesley, Inc., Menlo Park, California, 1990.

\bibitem{Hurd88phd}
L.~Hurd, {\em The application of formal language theory to the dynamical
  behavior of cellular automata}, Ph.D. thesis, Princeton University, Dept. of
  Math., June 1988.

\bibitem{Hurd90b}
\bysame, {\em Nonrecursive cellular automata invariant sets}, Complex Systems,
  1990.

\bibitem{Hurd90a}
\bysame, {\em Recursive cellular automata invariant sets}, Complex Systems {\bf
  4, No.2} (1990).

\bibitem{CulikHurd89a}
K.~Culik II, L.~P. Hurd, and S.~Yu, {\em Computation theoretic aspects of
  cellular automata}, In Gutowitz \cite{CALosAlamos89}.

\bibitem{CulikHurd89b}
\bysame, {\em Formal languages and global cellular automaton behavior}, In
  Gutowitz \cite{CALosAlamos89}.

\bibitem{Kurka92pre}
P.~Kurka, {\em Simulation in dynamical systems and {Turing} machines},
  preprint, Charles University, Dept. of Math. and Phys., Prague, 1992.

\bibitem{Kurka93pre}
\bysame, {\em Dynamical systems and factors of finite automata}, preprint,
  Charles University, Dept. of Math. and Phys., Prague, 1993.

\bibitem{Salomaa73}
A.~Salomaa, {\em Formal languages}, Academic Press, New York, 1973.

\bibitem{TrollSmil89}
G.~Trol{\noopsort{k}}l and U.~Smilansky, {\em A simple model for chaotic
  scattering, {I.~C}lassical theory}, Physica D {\bf 35} (1989), 34--64.

\bibitem{Troll91phd}
G.~Troll, {\em Transition to chaos in a simple scattering model}, Ph.D. thesis,
  Weizmann Institute, Rehovot 76100, Israel, September 1991.

\bibitem{Troll92pre}
\bysame, {\em Formal language characterization of transitions to chaos of
  truncated horseshoes}, preprint~13, SFB 288, TU Berlin, May 1992.

\bibitem{Troll92}
\bysame, {\em How to escape a sawtooth -- transition to chaos in a simple
  scattering model}, Nonlinearity {\bf 5} (1992), 1151--1192.

\bibitem{Walters82}
P.~Walters, {\em An introduction to ergodic theory}, Springer Verlag, New York,
  1982.

\end{thebibliography}
\end{document}